\documentclass[pre,aps,twocolumn,showpacs]{revtex4}
\usepackage{epsfig}
\include{graphics}
\begin{document}

\title{Two-level modeling of quarantine}

\author{Evgeniy Khain}
\affiliation{Department of Physics, Oakland University, Rochester, MI 48309, USA}

\begin{abstract}
Continuum models of epidemics do not take into account the underlying microscopic network structure of social connections. This drawback becomes extreme during quarantine when most people dramatically decrease their number of social interactions, while others (like cashiers in grocery stores) continue maintaining hundreds of contacts per day. We formulate a two-level model of quarantine. On a microscopic level, we model a single neighborhood assuming a star-network structure. On a mesoscopic level, the neighborhoods are placed on a two-dimensional lattice with nearest neighbors interactions. The modeling results are compared with the  COVID-19 data for several counties in Michigan (USA) and the phase diagram of parameters is identified.
\end{abstract}


\maketitle

\section{Introduction}

Reaction diffusion dynamics on a lattice is an active topic of current research \cite{lattice}. The intrinsic stochasticity significantly affects various macroscopic phenomena such as front propagation \cite{front} or phase transitions \cite{phase} and leads to completely new effects such as extinction in metapopulation models \cite{extinction}. One particularly interesting area of research deals with reaction diffusion dynamics on networks \cite{reactiondiffusion}.

The topic of the spread of epidemics on networks has received substantial attention in recent years \cite{review}. These network models eliminate the two main drawbacks \cite{Newman} of the standard SIR and SEIR models \cite{Murray} of the spread of an epidemic. The first drawback is related to the rate of recovery of an infected individual. The modeling implies a Poisson process, which means an exponential distribution of individual disease duration. This is in contrast to observations showing that the distribution is peaked around an average disease duration. The second drawback is the assumption of an equal number of contacts for each individual, i.e. ignoring the underlying microscopic structure of the social network \cite{Newman}.

Different individuals have a different average number of contacts, depending not only on their social behavior, but on their work. The inhomogeneity in the number of contacts becomes especially well-pronounced during the time of quarantine, when the majority of people work from home, but some individuals (like cashiers in a grocery store) still maintain hundreds of contacts per day. Typically, metapopulation models assume that the local neighborhoods are well mixed; then to model the entire population, these neighborhoods are placed on a lattice or form a network. The disease dynamics on such a network can be investigated by taking into account various migration patterns of individuals between the neighborhoods \cite{travel}. The present work formulates a basic model of disease dynamics during the quarantine, testing the other extreme, where each single neighborhood is far from being well-mixed and is modeled by a star-like network, while a larger region is modeled as a lattice of these neighborhoods.

\section{The model}
The model consists of two levels. The microscopic modeling describes a single neighborhood assuming a star network \cite{star}, a structure, where every node (a household) is connected to the central hub (a grocery store). The neighborhood consists of a large number of households (denoted by $N$) not connected to each other and not interacting with each other (mimicking the quarantine). A representative from each household visits the grocery store twice a week and interacts with a cashier. If the store visitor is ill, the cashier can be infected with probability $\beta$ or vice-versa: if the cashier is ill, the  store visitor can be infected with probability $\beta$ (this is the first important parameter of the model). Different stages of the disease are considered. Apart from susceptible individuals (healthy individuals who can catch the disease), there are exposed individuals who are infected but cannot infect others (this period lasts approximately $5$ days \cite{incubation}). The next step in the disease progression is being infected without symptoms; it is assumed that this period lasts $3$ days and during this time the infected individual can infect others. Then a person might develop symptoms, in which case they stay at home (this period lasts about $15$ days). Finally, an infected individual can recover or die. The overall considered duration of the disease for a single individual is in agreement with the literature \cite{duration}.

The main idea of the star network is the importance of the central node: the cashier. Once a cashier gets the infection and shows symptoms, they are replaced by a new cashier. During the time the cashier is already infected and still working, they can infect many customers, leading to an outbreak of the disease in the neighborhood. Monte-Carlo simulations show that in the majority of cases, the infected customers eventually infect the new cashier, continuing the outbreak. Prescribing a certain mortality rate (the second important parameter of the model), one can compute the average number of deaths in a single neighborhood as a function of time since the start of the epidemic.

The duration of the outbreak in a single neighborhood (denoted by $\tau$) increases with $\beta$, but it is substantially shorter than the duration of the epidemic in a large county, containing hundreds of neighborhoods. To compute $\tau$, we performed $10000$ simulations of a single neighborhood, in which we produced an outbreak in a neighborhood by starting with an infected cashier. Each particular realization has its own number of infected cashiers, which translates to the duration of the outbreak: once the new cashier is not infected, the epidemic dies out in a star network. The chance that a large number of cashiers are infected one after another is exponentially small, so averaging over many simulations gives an exponential decrease in the number of exposed individuals in a neighborhood. This time dependence of the average number of exposed individuals was measured in simulations, and we performed an exponential fit in the form $A\exp(-t/\tau)$. Figure 1 shows these measurements and the resulting characteristic duration of the outbreak $\tau$ as a function of $\beta$ for two values of $N$ ($1000$ households and $700$ households in a neighborhood).

\begin{figure}[ht]
\begin{center}
\includegraphics[width=2.5 in]{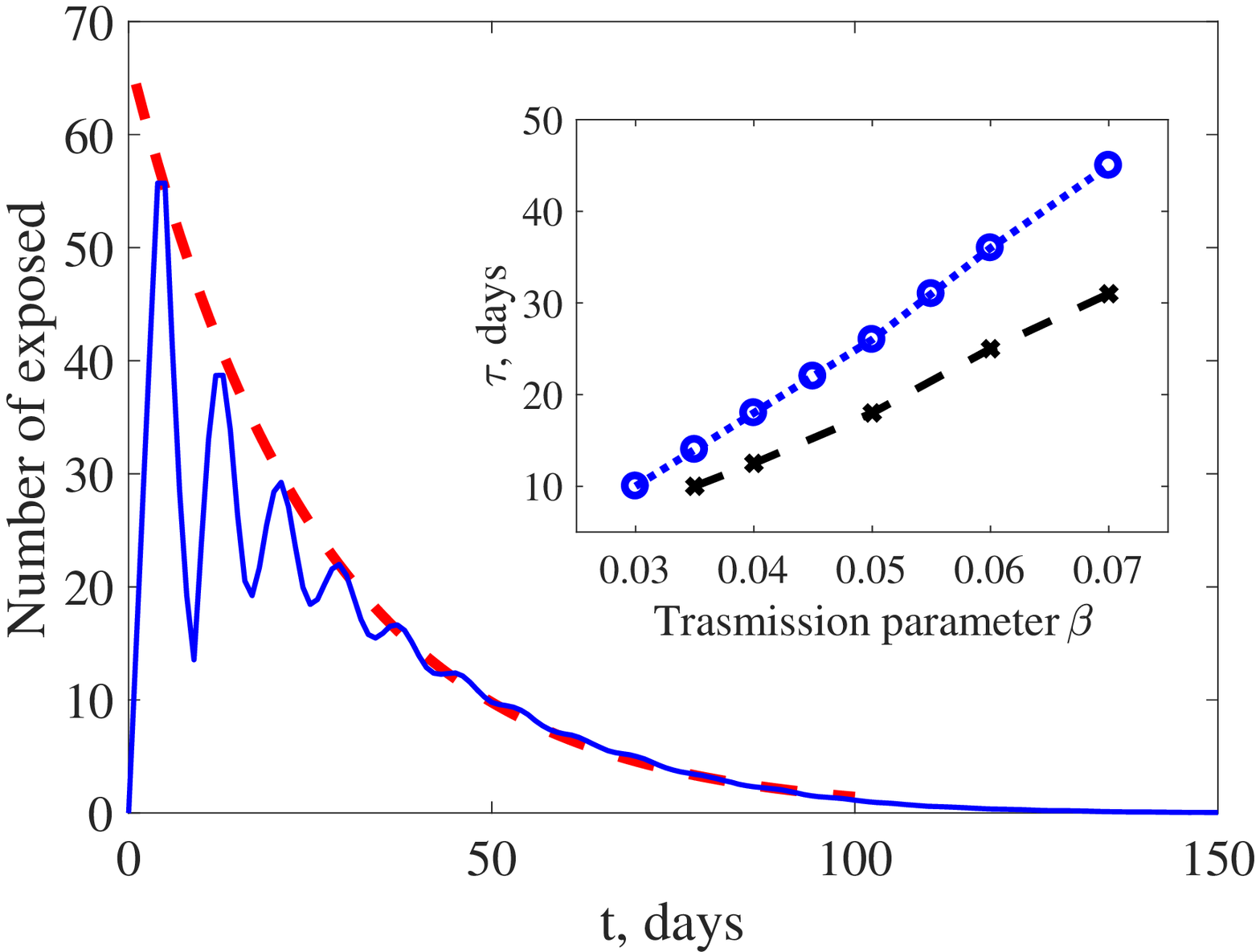}
\caption{The number of exposed individuals in a single neighborhood as a function of time for $\beta=0.05$. The exponential fit (red dashed curve) provides the characteristic duration of the outbreak $\tau$. The inset shows this $\tau$ as a function of $\beta$ for two values of the number of households in a neighborhood: $N=1000$ (the dashed line with circles) and $N=700$ (the dotted line with x symbols).
\label{fig:diseasetime}
}
\end{center}
\end{figure}

As expected, the outbreak in a single neighborhood lasts longer for higher values of $\beta$ and for larger number of households $N$: for the higher $\beta$, it is easier to infect a new cashier, prolonging the outbreak.  Notice that the characteristic duration of the outbreak in a neighborhood, $\tau$, is not the same as the individual illness time. This duration time $\tau$ (or the rate of recovery of a neighborhood, $1/\tau$) is used in the next level of modeling: many neighborhoods on a lattice, see Figure 2 for the schematic representation of the system.

\begin{figure}[ht]
\begin{center}
\includegraphics[width=2.5 in]{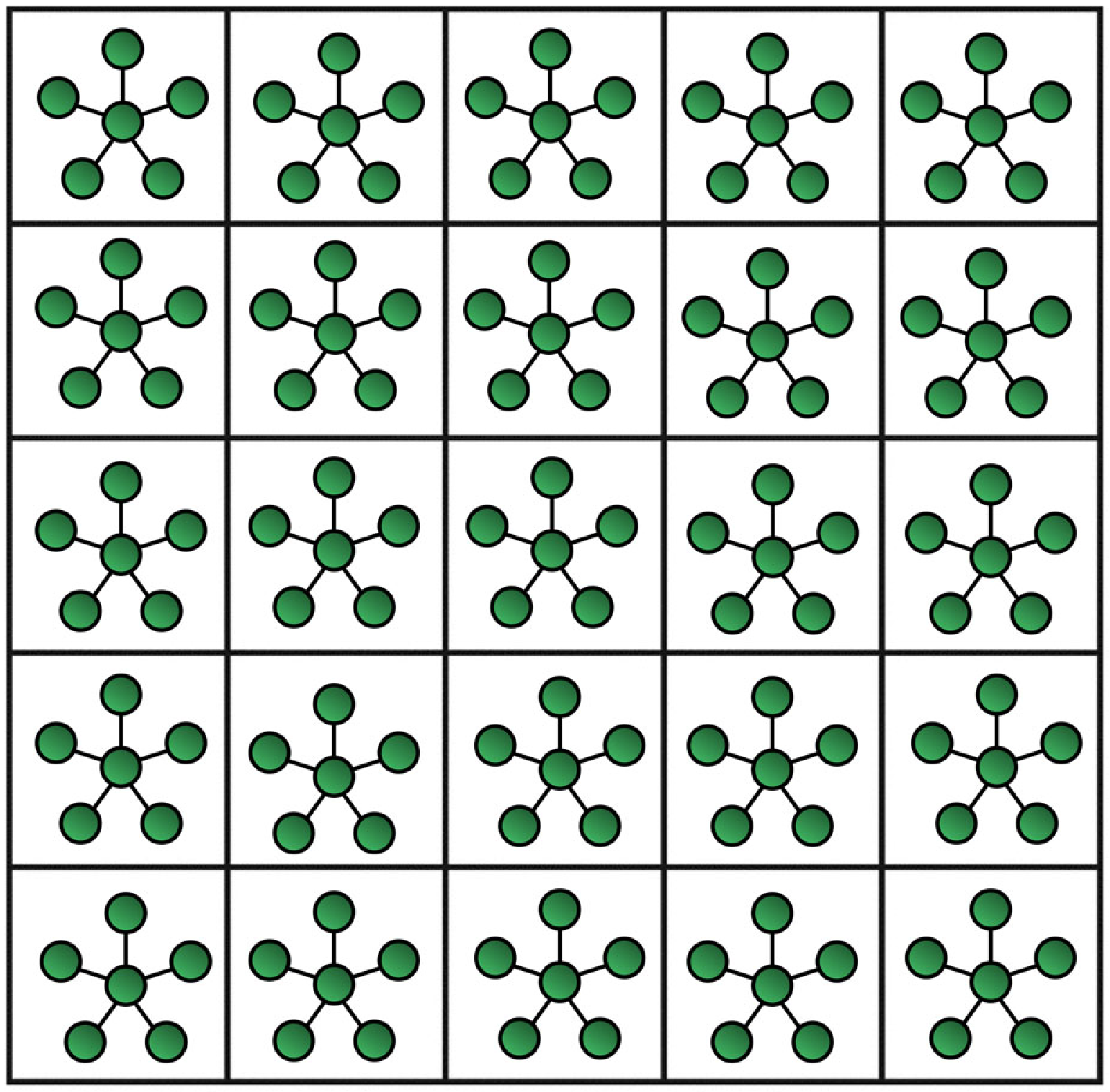}
\caption{Schematic representation of the system: weakly coupled SIR-like neighborhoods on a lattice. Each neighborhood has a star-network structure.
\label{fig:schematic}
}
\end{center}
\end{figure}

When on a lattice, each neighborhood can be in one of three states: susceptible, infected or recovered. Initially, all of the neighborhoods are susceptible, but since some households are already infected, there is a certain initial rate of ``self-infection" of a susceptible neighborhood. The neighboring lattice sites are weakly interacting. There are no interactions between the usual people (the leaf nodes) from different neighborhoods. Instead, we assume that a representative from each household from one neighborhood visits grocery stores in the neighboring neighborhood, but these visits (once per month to each of the four neighboring grocery stores) are significantly less frequent than the visits to their own grocery store (twice per week). Still, a susceptible neighborhood can catch the infection from a neighboring infected neighborhood. All of the relevant rates are measured in the ``microscopic" single neighborhood simulations. Then we performed Monte-Carlo simulations of neighborhoods on a lattice and, measuring the times at which various neighborhoods got infected, we computed the overall death toll in the county and the number of cases as a function of time. The consensus in the community is that the official total number of cases is substantially underestimated, since only a fraction of infected people is tested, so instead of relying on the reported number of cases, we compared our results with the Michigan death toll data. Figure 3 shows this comparison for Oakland County, the second largest county in Michigan with a population of over $1.2$ million people. One can see that a perfect agreement is achieved for two different sets of parameters: low $\beta$ (less infectious) with high mortality and high $\beta$ (more infectious) with low mortality.

\begin{figure}[ht]
\begin{center}
\includegraphics[width=3.0 in]{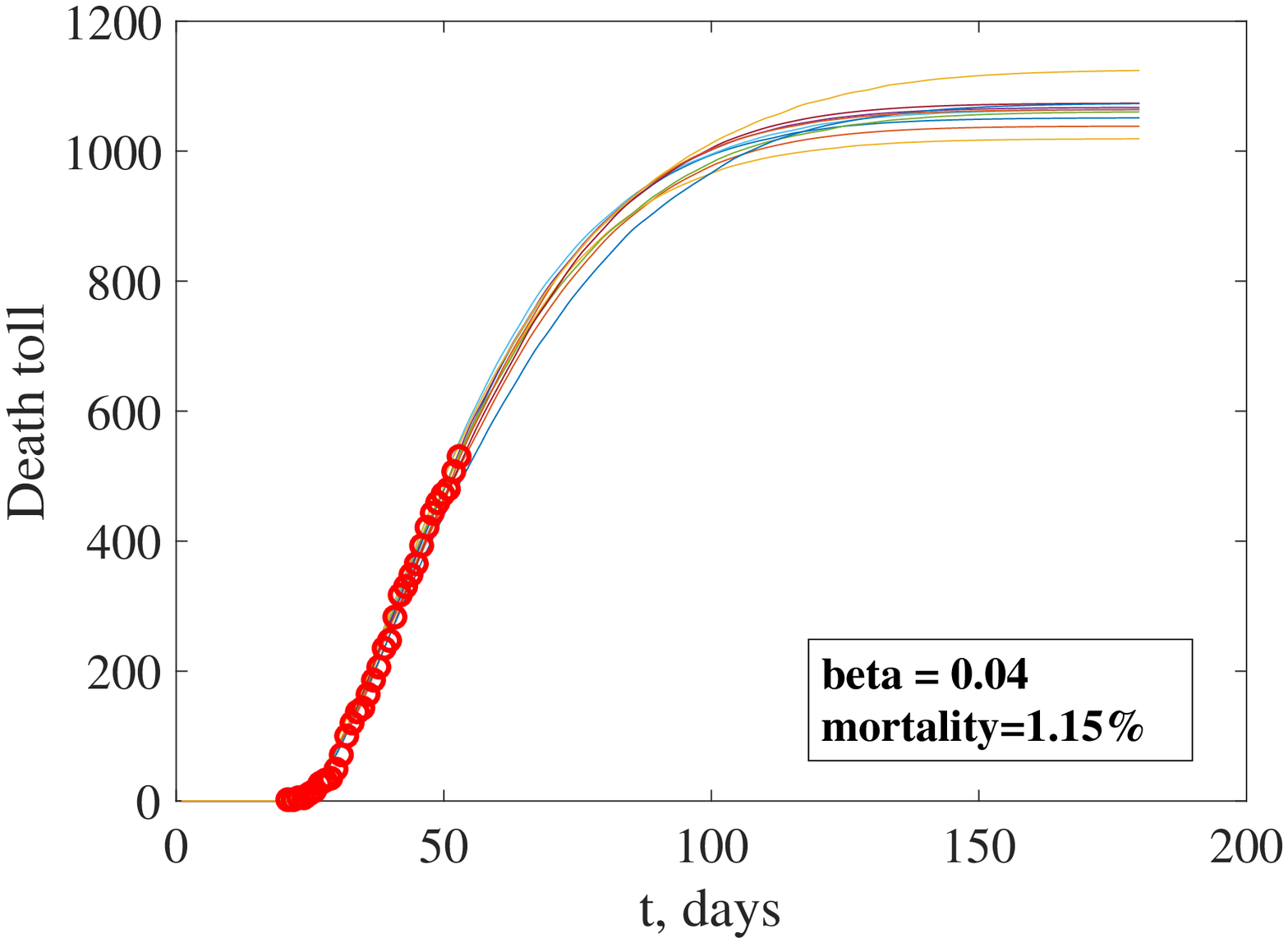}
\includegraphics[width=3.0 in]{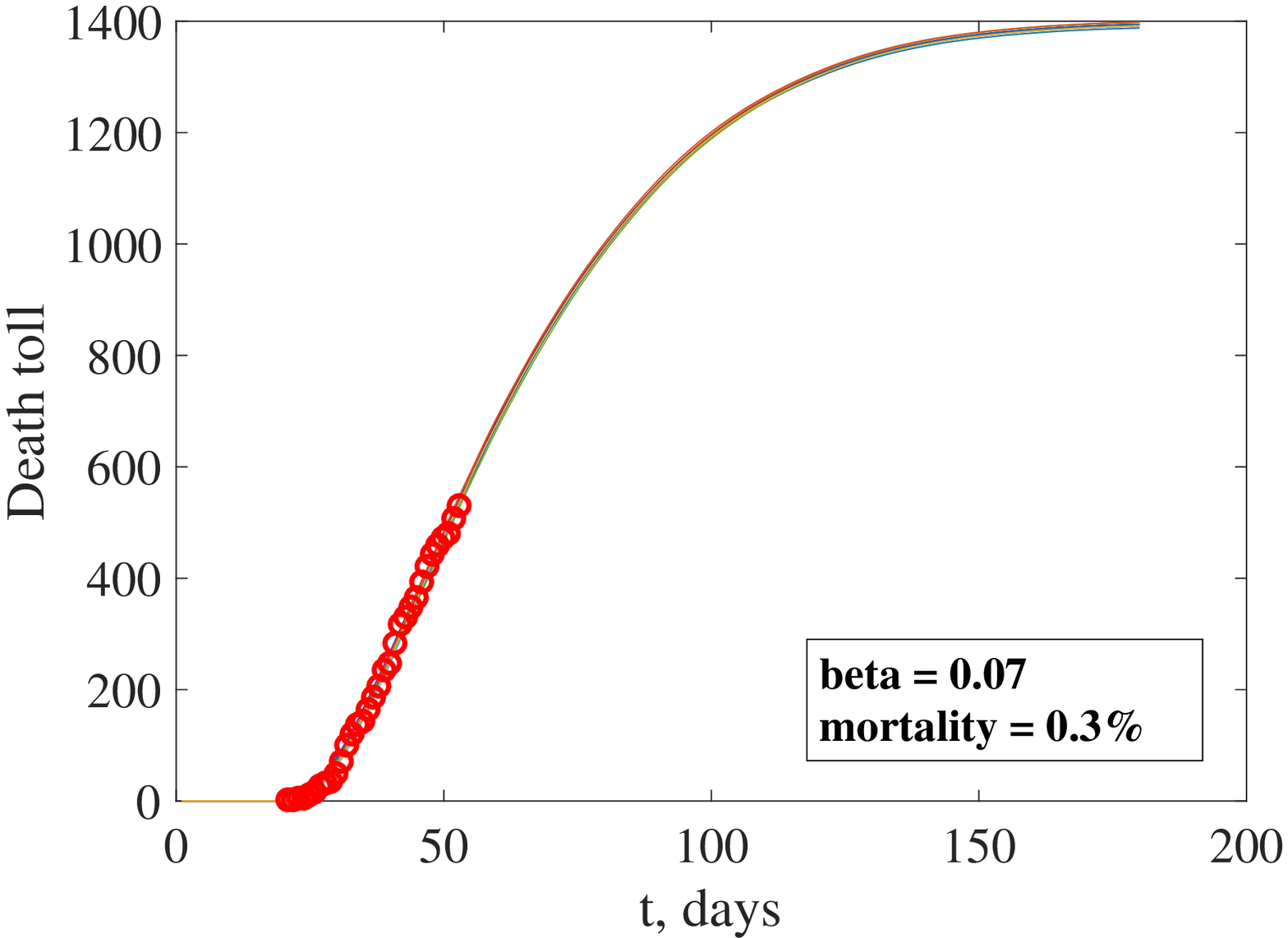}
\caption{Death toll as a function of time in Oakland county, Michigan (USA). The initial time is March 1, 2020. In each panel, circles represent the official data \cite{Michigan}, and solid curves show the results of simulations of neighborhoods on a lattice ($10$ such simulations are shown in each panel to demonstrate the effect of stochasticity). The upper panel corresponds to $\beta=0.04$ and mortality of $1.15$ percent, while the lower panel corresponds to $\beta=0.07$ and mortality of $0.3$ percent. Both panels show an excellent agreement with the data. Assuming $N=1000$ households in a neighborhood with an average of $3$ persons in a household, Oakland county was simulated on a $20$x$20$  lattice, a total of $400$ neighborhoods.
\label{fig:oakland}
}
\end{center}
\end{figure}

\begin{figure}[ht]
\begin{center}
\includegraphics[width=3.8 in]{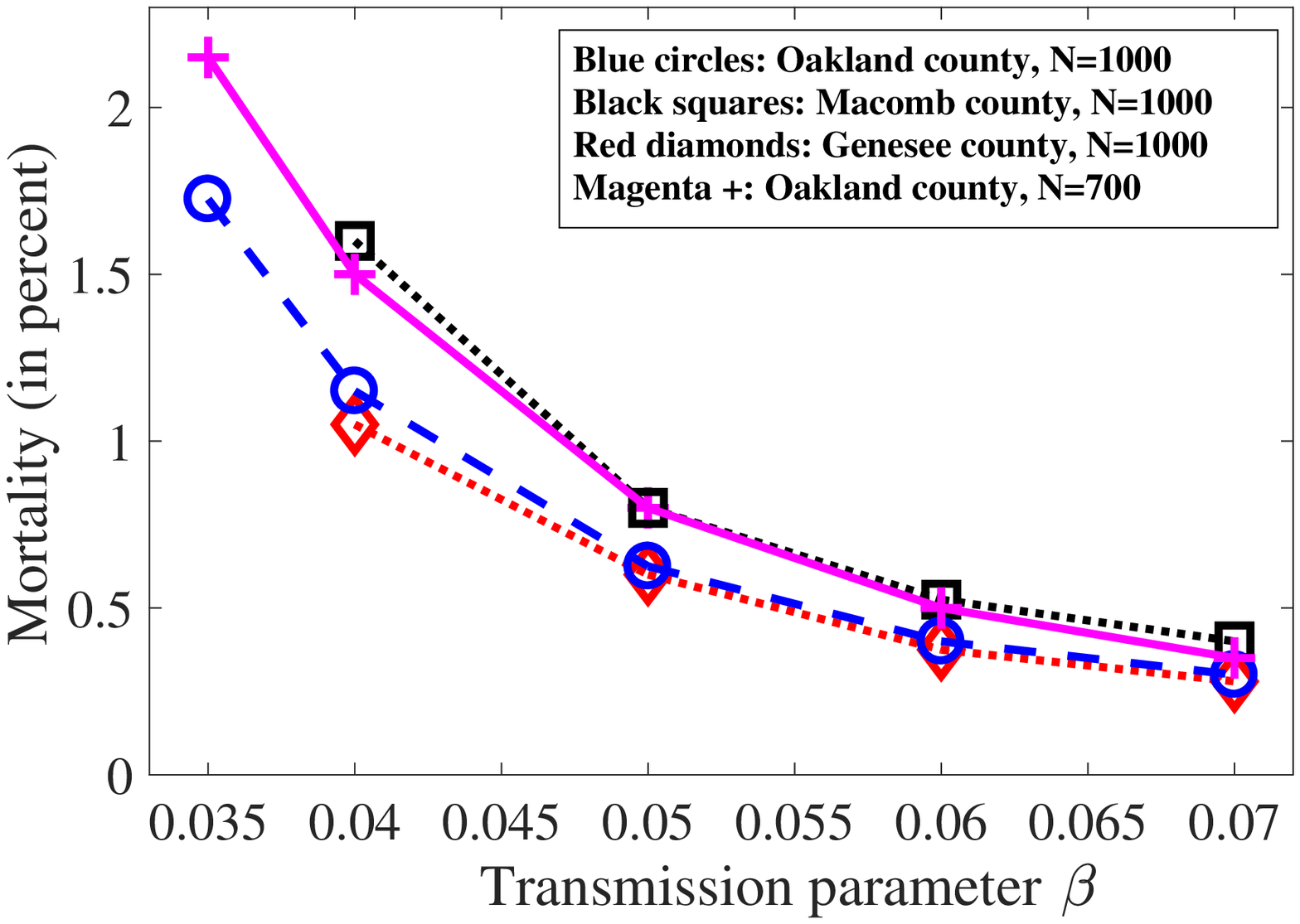}
\caption{Phase diagram of parameters for three counties in Michigan (USA). Each point of the curve corresponds to a set of parameters that perfectly describes the current death toll data for the respective county. Assuming $N=1000$, Oakland county (dashed line, circles) was simulated on a $20$x$20$ lattice, Macomb county (dotted line, squares) was simulated on a $17$x$17$ lattice, and Genesee county (dotted line, diamonds) was simulated on a $12$x$12$ lattice. For comparison, we also show simulations of Oakland county for $N=700$ and $24$x$24$ lattice (solid line, pluses).
\label{fig:phasediagram}
}
\end{center}
\end{figure}

This degeneracy implies that there is a curve on the phase plane of parameters ($\beta$-mortality), such that each point on this curve describes the current death toll data (as of April $20$, $2020$) well. We computed this curve not only for Oakland county but also for two other counties in Michigan: Macomb county (with total population of above $850$ thousand) and Genesee county (with total population above of $400$ thousand), see Figure 4. We have also checked the effect of the size of neighborhoods. The total population of Oakland county is known and fixed in the modeling. Therefore, the size of the neighborhood fully determines the lattice size. This is exactly what Figure 4 shows for Oakland county: smaller neighborhoods on a larger lattice (solid line, pluses) and larger neighborhoods on a smaller lattice (dashed line, circles). All the parameters in the figure fit the data very accurately, so there is a (small) uncertainty in determining the mortality even for a fixed value of $\beta$; the uncertainty is related to the fact that there can be smaller and bigger neighborhoods. The reason for this effect is that for the same $\beta$, larger neighborhoods have stronger outbreaks. Therefore, a smaller mortality parameter is required to obtain the observed death toll.

The main question now is how one can constrain the parameter space. Which set of parameters ($\beta$, mortality) is more reasonable? First, one would like to describe these counties with similar values of $\beta$ and mortality rates. Therefore, the top left corner of the phase space is not a good region to try since the curves move apart. Another reason for not choosing parameters in the top left region of the diagram is that for low contagiousness (low $\beta$) and high mortality, the epidemic is almost over, and unfortunately, we are not there yet. The low right region of the phase space is not a good candidate either. For a highly contagious disease (high $\beta$), a substantial fraction of the population is already infected. For example, simulations show that for $\beta=0.07$, more than one third of the total number of households in Oakland county would already be infected (as of April $24$, $2020$). This number is too high as can be seen from the testing data: less than $20-25$ percent of tests in Michigan are positive. Since this data is for people who are tested (people from a high risk group with some symptoms), this is clearly an upper bound for the fraction of currently infected individuals. As a result, the reasonable region of the phase diagram is in the middle, for example, with $\beta  \simeq  0.05$.

Choosing parameters from this (middle) region of the phase diagram, one can compute the fraction of susceptible households as a function of time, see the solid curve in Figure 5. If the quarantine is not lifted, approximately three quarters of the population will not catch the disease. The dotted line, however, shows the fraction of susceptible cashiers: most of them get infected during the epidemic. The observation that nodes with high degree (in our case, the cashiers) are much more likely to be infected is known in the literature as the ``$20/80$ rule" \cite{2080}. The inset shows the fraction of infected households, one can see that the peak was reached in the middle of April.

\begin{figure}[ht]
\begin{center}
\includegraphics[width=3.0 in]{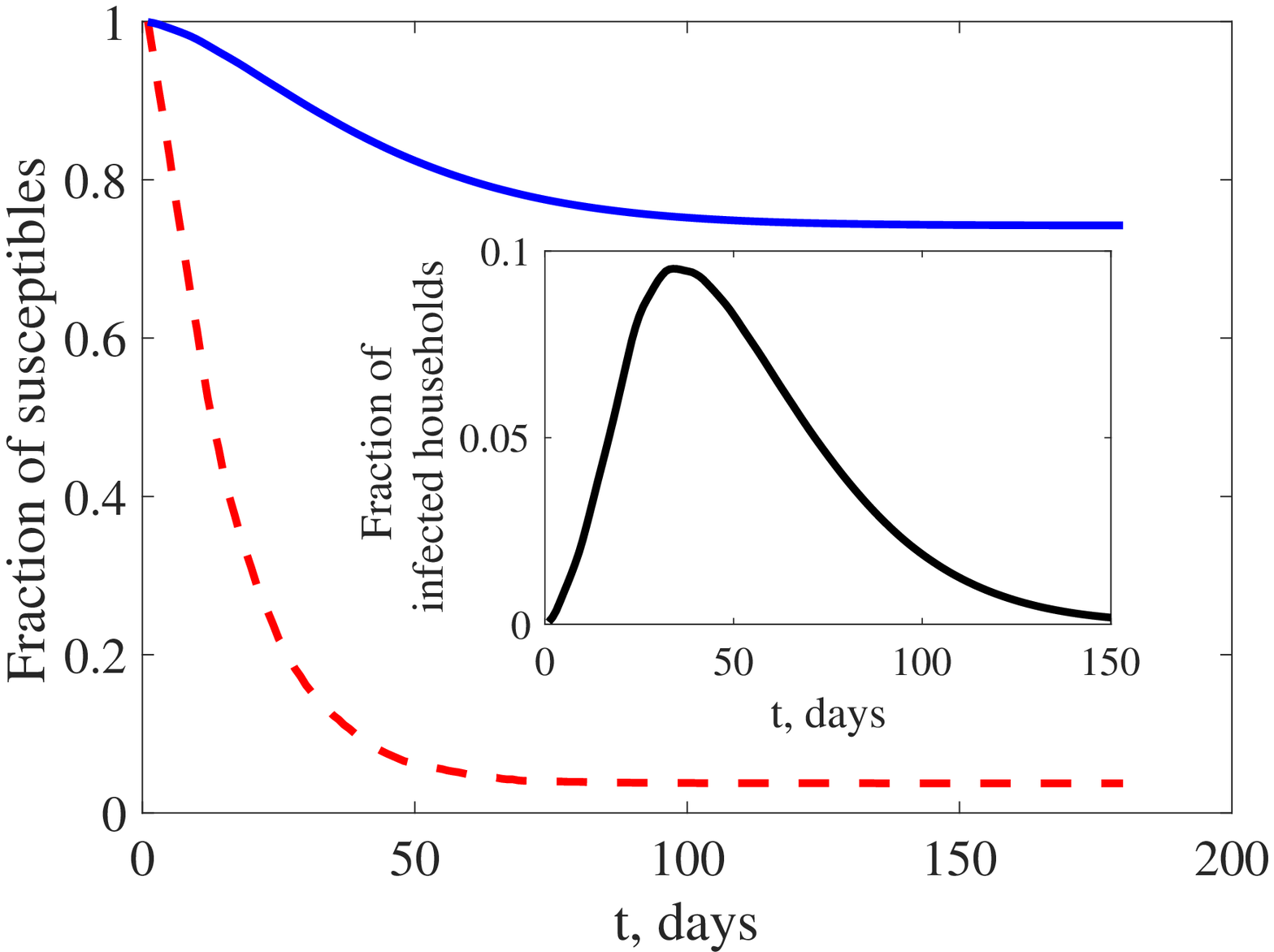}
\caption{Fraction of susceptible households (blue solid curve) and fraction of susceptible cashiers (red dashed curve) as a function of time. The inset shows the fraction of infected households. Simulations of Oakland county, Michigan on a $20$x$20$ lattice for $N=1000$, $\beta=0.05$ and mortality of $0.62$ percent.
\label{fig:sus}
}
\end{center}
\end{figure}

In the revised version of the paper, we were able to compare the results of our simulations with the death toll data until June $15$, $2020$. Since gatherings of up to $10$ people in the state of Michigan were allowed starting from May $21$ and the ``stay at home order'' was lifted on June $1$, the death toll data for a later period cannot be described by the current quarantine model with star-like networks for individual neighborhoods. As predicted by the model (see Figure 3), the death toll curve saturates. Figure 6 shows a nice agreement with the data; using the standard least square method, the best fit was obtained for $\beta$ = $0.039$ and a mortality of $1.29$ percent, which corresponds to the left region in the phase diagram (Figure 4). Figure 6 also shows that approximately $10.3$ percent of the population in Oakland county caught the disease, which is about $123600$ individuals. The reported number of cases is much lower: the official number of cases reported by June $15$ was $8564$ (and another $2749$ cases are in question) \cite{Michigan}. Therefore, the official data underestimates the number of cases by a factor of $11$ to $14$.

\begin{figure}[ht]
\begin{center}
\includegraphics[width=3.5 in]{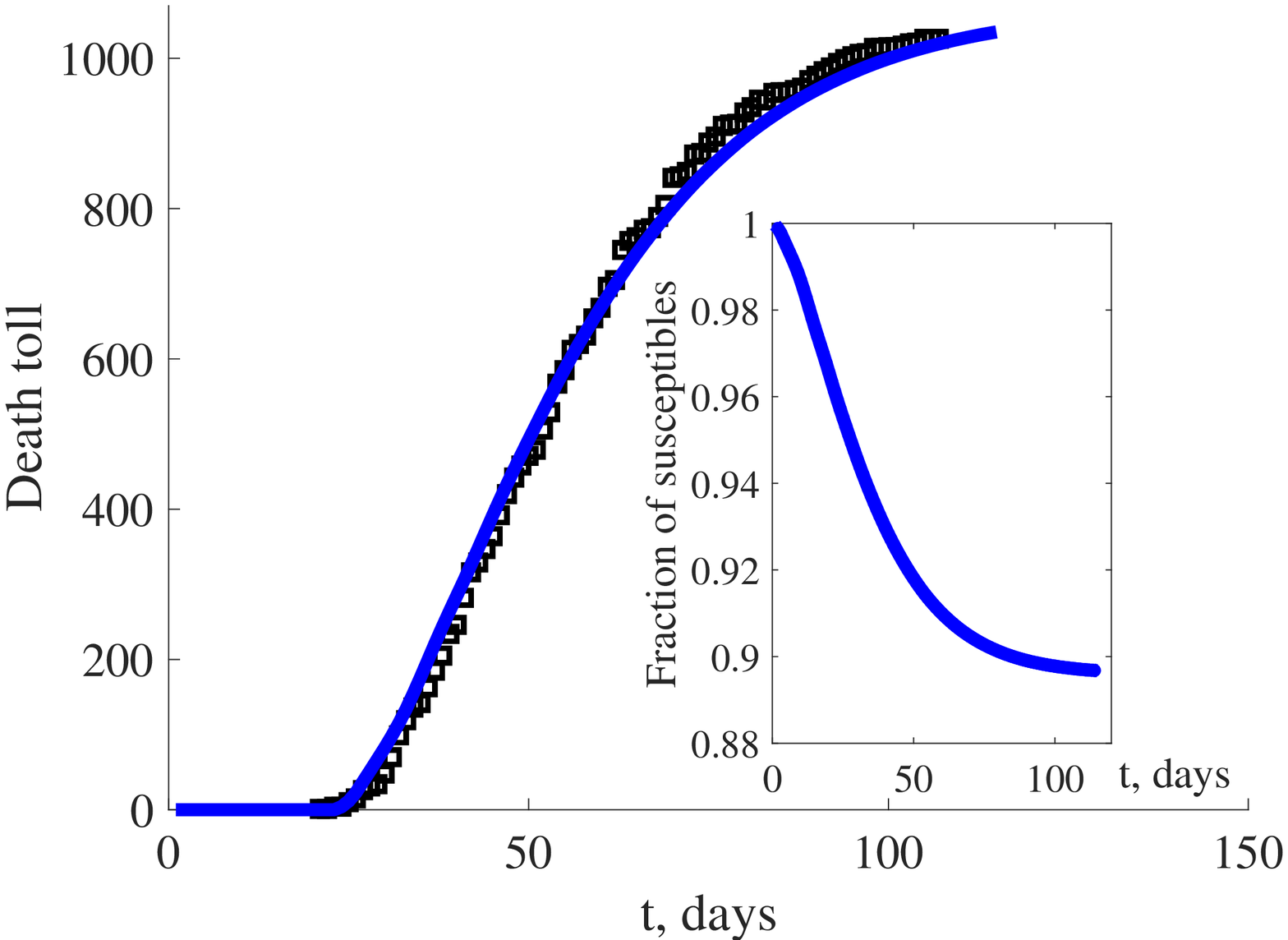}
\caption{Death toll as a function of time in Oakland county (Michigan): official data (black squares) and simulations of the model (blue solid line). This figure takes into account the official Michigan death toll data up to June $15$. The inset shows the fraction of susceptible individuals. Simulations of Oakland county are performed on a $20$x$20$ lattice for $N=1000$, $\beta=0.039$ and mortality of $1.29$ percent.
\label{fig:sus}
}
\end{center}
\end{figure}

\section{Summary and discussion}
This work focuses on modeling disease dynamics during the quarantine, when most people dramatically decrease their number of contacts, but some individuals still maintain hundreds of contacts per day. On a microscopic level, this pattern is modeled by a star network, where the central node (say, a cashier in a neighborhood) is connected to all other nodes, but all other connections are prohibited. A big county is modeled as many such neighborhoods on a lattice, Figure 2. Simulations show that the results do not strongly depend on the number of households $N$ in a single neighborhood if changing $N$ is compensated by adjusting the number of neighborhoods on a lattice to keep the county population constant. However, the results strongly depend on the two main parameters: the transmission coefficient $\beta$ and the mortality. Stochastic simulations of this two-level model show a sloppy behavior \cite{Sethna}: different sets of these parameters can describe the same death toll data in a county. We were able to identify the region in the phase plane of parameters that reproduces the observations in different counties and estimate the mortality and the infection probability. Analyzing the later data, we obtained the mortality in Oakland county to be around $1.3$ percent, which also suggests that the real number of coronavirus cases in this county is $11-14$ times larger than the number of reported cases. This number is $5$ times lower than in a recent controversial study in Santa Clara County \cite{Stanford}, but is still very high. In some places, where the outbreak is (was) particularly severe (for example, certain parts of Italy), the official death toll might be underreported \cite{Italy}. The situation in Michigan hospitals is substantially better, so it was assumed that the Michigan death toll data is accurate.

The presented model is phenomenological and aims at capturing basic features of a quarantine and avoiding the drawbacks of continuum modeling. It can be easily modified to include more details: other essential workers with a large number of connections (for example, health care workers and more cashiers in the same grocery store) and population structure in a county. Yet, phenomenological models are useful for our basic understanding of the underlying physical mechanisms and can produce good predictions. Figure 5 shows that the fraction of infected cashiers is very large, justifying the assumption that a recovered neighborhood during the quarantine can not be infected again. However, when the quarantine is lifted, more high-degree nodes (people with many  contacts per day, for example, university teachers of general physics classes) will return to work, a substantial fraction of whom are still susceptible. This is likely to lead to the second wave of the disease outbreak.

In order to perform simulations of neighborhoods on a lattice, one needs to use a microscopic model of a neighborhood to compute the recovery rate. This idea of two-level modeling, when the rates are measured in a microscopic model and then used in on a macroscopic level, has recently been employed in a completely different problem of rare cell clustering on a substrate \cite{KhainKhasin}.

\begin{acknowledgments}
The author thanks Misha Khasin, Oleg Kogan and Roger from ChessPro forum for fruitful discussions. The author is also in debt to Tali Khain, Daniel Khain and Julia Falkovitch-Khain for many extensive discussions during the quarantine.
\end{acknowledgments}


\end{document}